\documentclass[11pt]{article}
\usepackage{blois,epsfig}

\def\be{\begin{equation}}
\def\ee{\end{equation}}
\def\bea{\begin{eqnarray}}
\def\eea{\end{eqnarray}}

\begin{document}
\vspace*{2cm}
\title{SELECTED RESULTS FROM THE ANTARES NEUTRINO TELESCOPE}

\author{ S. MANGANO, FOR THE ANTARES COLLABORATION}

\address{IFIC - Instituto de F\'isica Corpuscular, Edificio Institutos de Invesstigaci\'on, \\
Apartado de Correos 22085, 46071, Spain}

\maketitle\abstracts{
The ANTARES telescope is the largest underwater neutrino telescope existing at present.
It is based on the detection of Cherenkov light produced in sea water by neutrino-induced muons.
The detector, consisting of a tri-dimensional array of 885 photomultipliers arranged on twelve vertical lines, is
located at a depth of 2475 m in the Mediterranean Sea, 40 km off the French coast.
The main goal of the experiment is to probe the Universe by means of neutrino events
in an attempt to investigate the nature of high energy astrophysical sources, to contribute
to the identification of cosmic ray sources, and to explore the nature of dark matter.
In this contribution we will review the status of the detector, illustrate its operation and performance,
and present the first results from the analysis carried out on atmospheric muons and neutrinos, as well as
from the search for astrophysical neutrino sources.}

\section{Introduction}
The main goal of the ANTARES neutrino telescope experiment is the observation
of cosmic neutrinos. The advantage of using neutrinos with respect to other cosmic particles are that they 
are not deflected by magnetic fields and are weakly interacting. The neutrinos point directly back to the source of emission and can provide
an unbiased information about the source.

ANTARES is located in the 
Mediterranean Sea, \mbox{40 km} off the French coast at \mbox{$42^{\circ} 50'$N, $6^{\circ} 10'$E}.
The detector was completed in May 2008 and consists of twelve vertical
lines equipped with 885 photomultipliers in total, installed at a depth of 2475 m. Before completion ANTARES has been taking data in different
line configurations.
The distance between adjacent lines is of the order of 70 m. 
Each line is equipped with
up to 25 triplets of photomultipliers spaced vertically by 14.5 m. The
photomultipliers are oriented with their axis pointing at  $45^{\circ}$ from the
vertical. The instrumented detector volume is about 0.02 $\textrm{km}^3$.

The detection principle relies
on the observation of
Cerenkov light emitted by relativistic charged particles in water and is optimized for the detection of muons. 
The muon trajectory is reconstructed from the arrival times of the Cerenkov light
detected by the photomultipliers, whose positions are monitored by means of a positioning system.

\section{Selected results: search for cosmic neutrinos}
A variety of
analyses have been performed since the data taking started with the
installation of the first line in 2006. In this paper we will focus on
three different data analyses which are related to the search for cosmic
neutrinos. There is not enough room to discuss here also of the other
topics which were illustrated at the workshop, such as the atmospheric
muons and the spatial and temporal correlation of neutrinos with cosmic
messengers like photons, cosmic rays and gravitational waves, for which
the reader is referred to elsewhere~\cite{ag}~\cite{ba}~\cite{pr}.

\begin{figure}
 \vspace*{-3cm}
 \setlength{\unitlength}{1cm}
 \centering
 \begin{picture}(18.5,7.5)
\put(-0.5,0.0){\epsfig{file=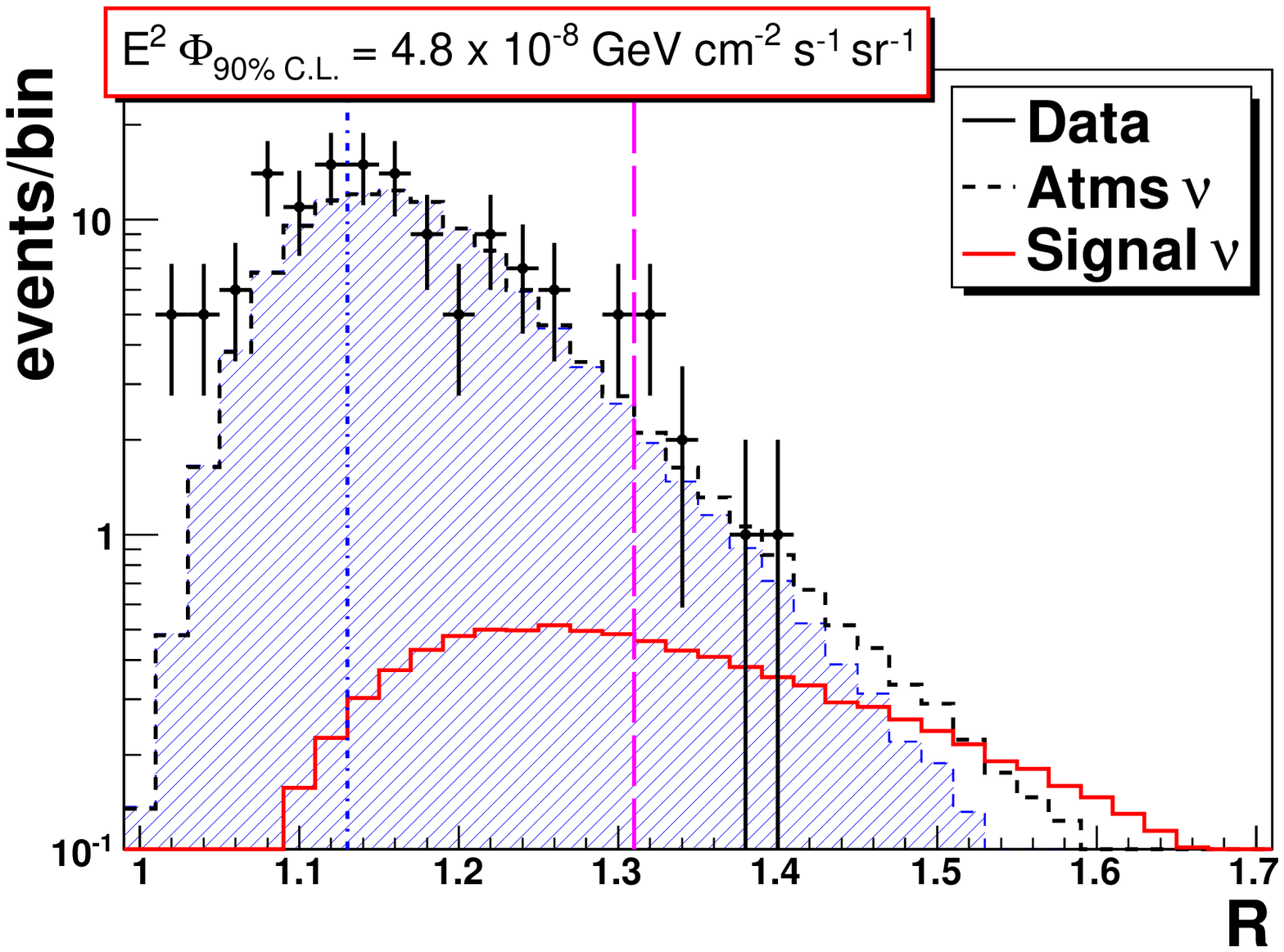,width=8.0cm,clip=}}
\put(7.3,-0.8){\epsfig{file=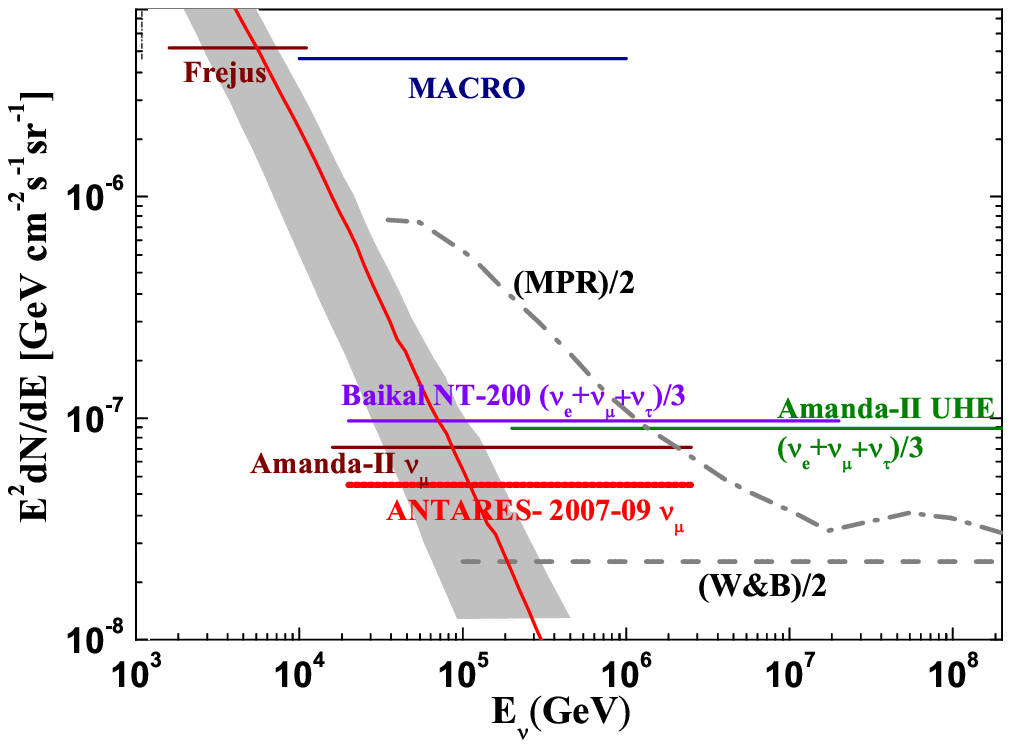,width=8.5cm,clip=}}
 \end{picture}
\caption[Sl]{Left: The distribution of the energy estimator parameter for reconstructed events (points), simulated events from a flux of atmospheric neutrinos normalized to the data (dashed line) and from a signal neutrino distribution normalized to the upper limit given in top of the Figure (solid line). Right: The upper limit (\mbox{90\% C.L.}) for a $E^{-2}$ diffuse muon neutrino flux compared to the limits from other experiments (Frejus~\cite{Frejus}, Macro~\cite{MACRO}, \mbox{Amanda-II}~\cite{Amanda_numu}, Amanda-II UHE~\cite{Amanda-UHE} and Baikal~\cite{Baikal}) and to predictions (W\&B~\cite{wb}and MPR~\cite{mpr}).}
\label{fig:diffflux}
\end{figure}


The main physical background to identify cosmic neutrinos are atmospheric muons and upward going atmospheric neutrinos.
A significant fraction of the atmospheric muons which are produced in the upper atmosphere by the interaction of cosmic rays can reach the apparatus despite the shielding
provided by 2 km of water. At the depth of ANTARES, the flux of atmospheric muons is around $10^6$ times higher than that of atmospheric neutrinos.
The muons produced by the interaction of neutrinos can be isolated from the atmospheric muons by selecting upward going particles. 
Upward going atmospheric neutrinos will be also detected in this way. The way to search for a signal of cosmic neutrinos is to look at their direction or to try to discriminate the atmospheric neutrinos based on an estimator of the particle energy. 

\subsection{Diffuse cosmic neutrino flux}
\label{diffflux}

The diffuse cosmic neutrino flux analysis does neither use the time nor the location 
of the reconstructed neutrinos candidates in contrast to the point source search and dark matter
search presented in the following sections. The data used for this analysis were collected 
from \mbox{Dec. 2007} to Dec. 2009 with 334 days of livetime.

The diffuse flux analysis is based on the observation of an excess of high energy neutrinos above the 
irreducible background of the upgoing atmospheric neutrinos.
The challenge of this analysis is the correct evaluation of 
the background, especially the one represented by atmospheric muons in large bundles. 
Downward going particles, wrongly reconstructed as upward going can have the same signature as high energy neutrinos.
To study this effect one year of livetime
of atmospheric muons has been simulated with the MUPAGE package~\cite{ca}, where parametrized muon bundles are 
generated directly at the detector level.

An energy estimator has been used to discriminate between the high energy neutrino signal
and the atmospheric neutrino background. The Model Rejection Potential technique~\cite{hi} has been used to optimize the cuts
on the energy estimator in order to find the strongest constraint on theoretical signal models. After a cut in the 
energy estimator variable (R $>$ 1.31) nine events are found in the data sample (see Figure \ref{fig:diffflux} left), while around ten atmospheric neutrino events (Bartol flux plus prompt contribution) were expected. The upper limit for a $E^{-2}$ diffuse flux of muon neutrinos at Earth with 90\% confidence level is given
in Figure \ref{fig:diffflux} right, extended over an energy range from 20 TeV to 2.5 PeV. This is currently the most stringent upper limit set worldwide.

\subsection{Point sources}
\label{point}
\begin{figure}
  \vspace*{-3cm}
 \setlength{\unitlength}{1cm}
 \centering
 \begin{picture}(18.5,7.5)
\put(-0.5,0.0){\epsfig{file=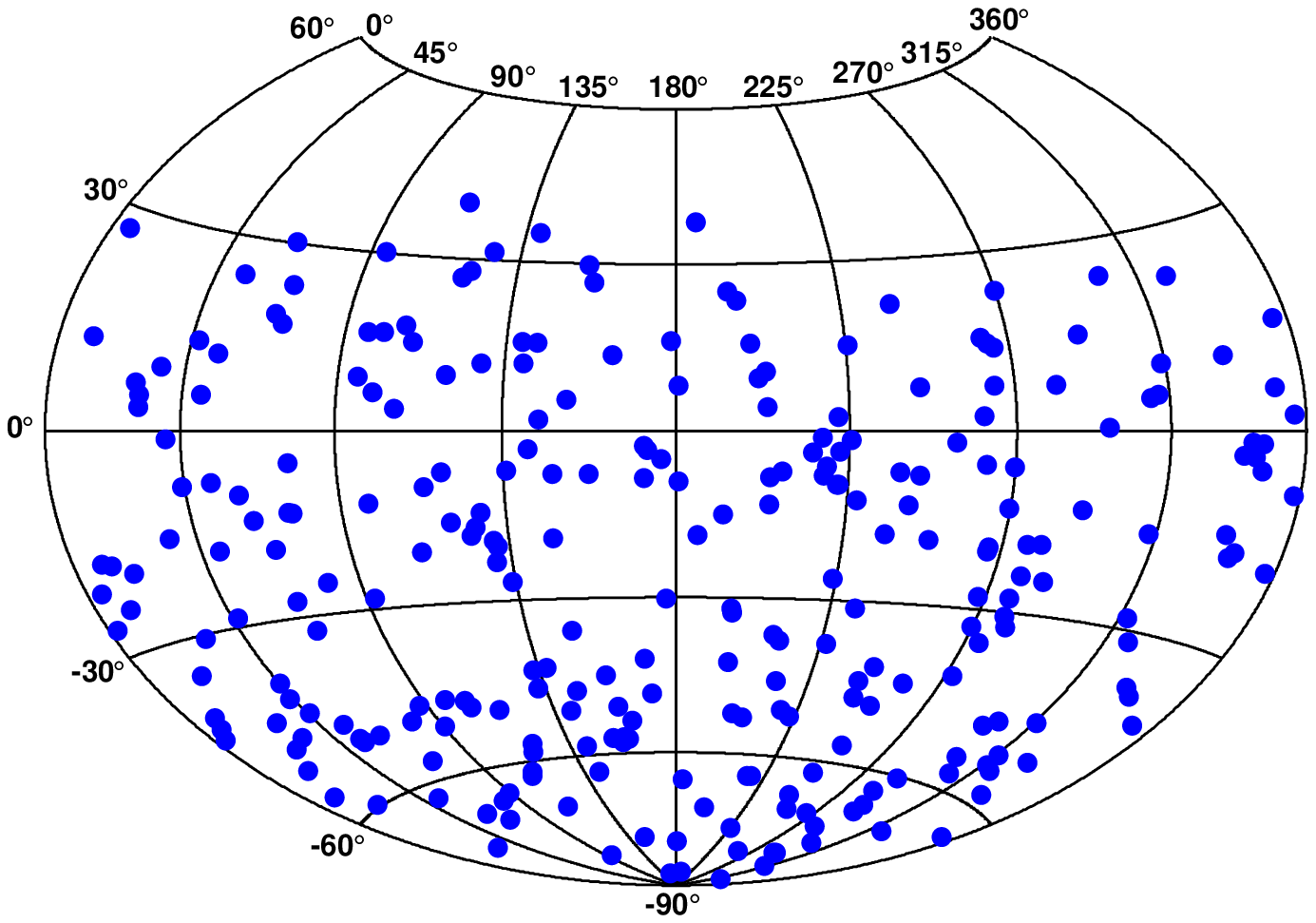,width=8.0cm,clip=}}
\put(7.5, -0.5){\epsfig{file=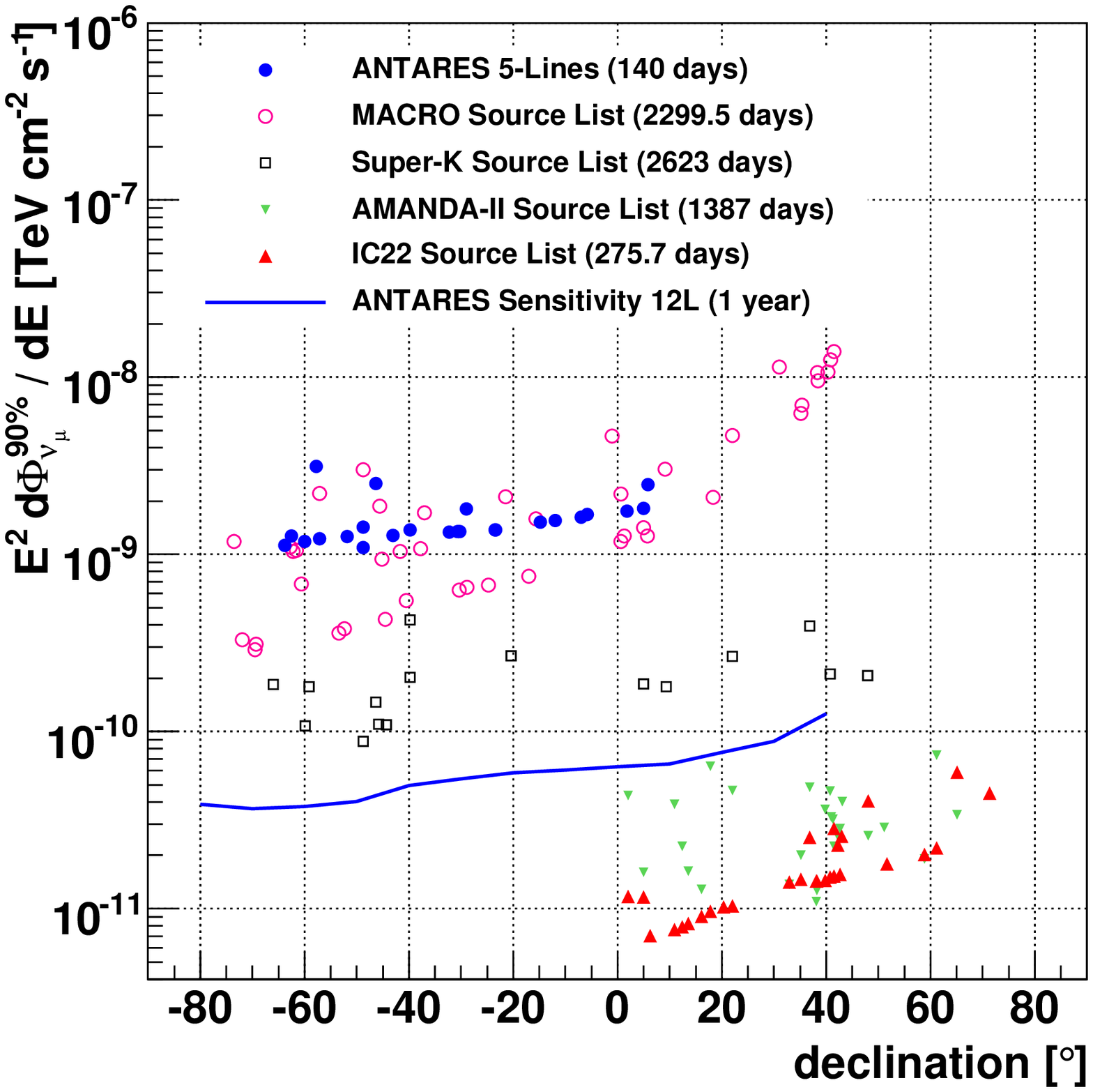,width=7.5cm,clip=}}
 \end{picture}
\caption[]{Left: Propagation directions of the 276 neutrino events selected from the 2007 data. The plot is in equatorial coordinates. Declinations above $47^{\circ}$ are always above the horizon and are invisible to the detector. Right: The upper point source limits for 2007 data (filled points) compared with the results published by other neutrino experiments (Macro~\cite{am}, Super-K~\cite{de}, Amanda~\cite{ab} and IceCube~\cite{ic}). The predicted sensitivity of ANTARES for 365 days (line) is also shown.}
\label{fig:point}
\end{figure}

Antares is in a very good location for the search of cosmic point sources, because
large part of the Galactic plain, where several sources of possible neutrino emitters exist, is visible.

A set of cuts have been optimized to search for an $E^{-2}$ neutrino flux in the data sample taken in 2007 extended over a livetime of 140 days. 
Taking the direction and time of detection of the 
276 upward going events which survive the cuts a sky view is obtained as shown in Figure \ref{fig:point} left. 
A fast and robust reconstruction algorithm 
with an angular resolution of around $3^{\circ}$ for energies above 10 TeV was used for this analysis. 
An all sky point source search based on an expectation-maximization method 
as well as a likelihood method did not reveal any significant excess for any direction.
As a further attempt, the information of the presence of galactic and extragalactic luminous source such as 
supernova remnants, pulsar wind nebulae and other gamma sources is used to constrain 
the search to specific regions
of the sky. By specifying the directions of the candidate sources the chance of the background to
imitate a signal is reduced. 24 sources have been selected among the most
promising neutrino source candidates. No significant
excess has been found. The Hess J1023-575 source with equatorial coordinates   
$\delta=-57.76^{\circ}, RA=155.8^{\circ}$ has the highest excess 
at the level of 1.8 $\sigma$. This is expected in 7.5\% of the experiments when looking at 24 sources. 
The upper limits on the neutrino flux from these 24 candidate sources are shown in Figure \ref{fig:point} right.
A more sophisticated reconstruction algorithm 
which can reach an angular resolution as low as $0.5^{\circ}$ above 10 TeV of particle energy is now being used in the analysis of the data collected in 2007 and 2008.

\subsection{Dark matter}
\label{sun}
\vspace*{-0.005cm}
The indirect search for dark matter in the universe is one further goal of ANTARES.
The relic neutralinos could gravitationally be trapped in the Sun and thereby increase the local 
neutralino density. In the subsequent neutralino annihilation process neutrinos can be created.

\begin{figure}
  \vspace*{-3cm}
 \setlength{\unitlength}{1cm}
 \centering
 \begin{picture}(18.5,7.5)
\put(-0.5,0.3){\epsfig{file=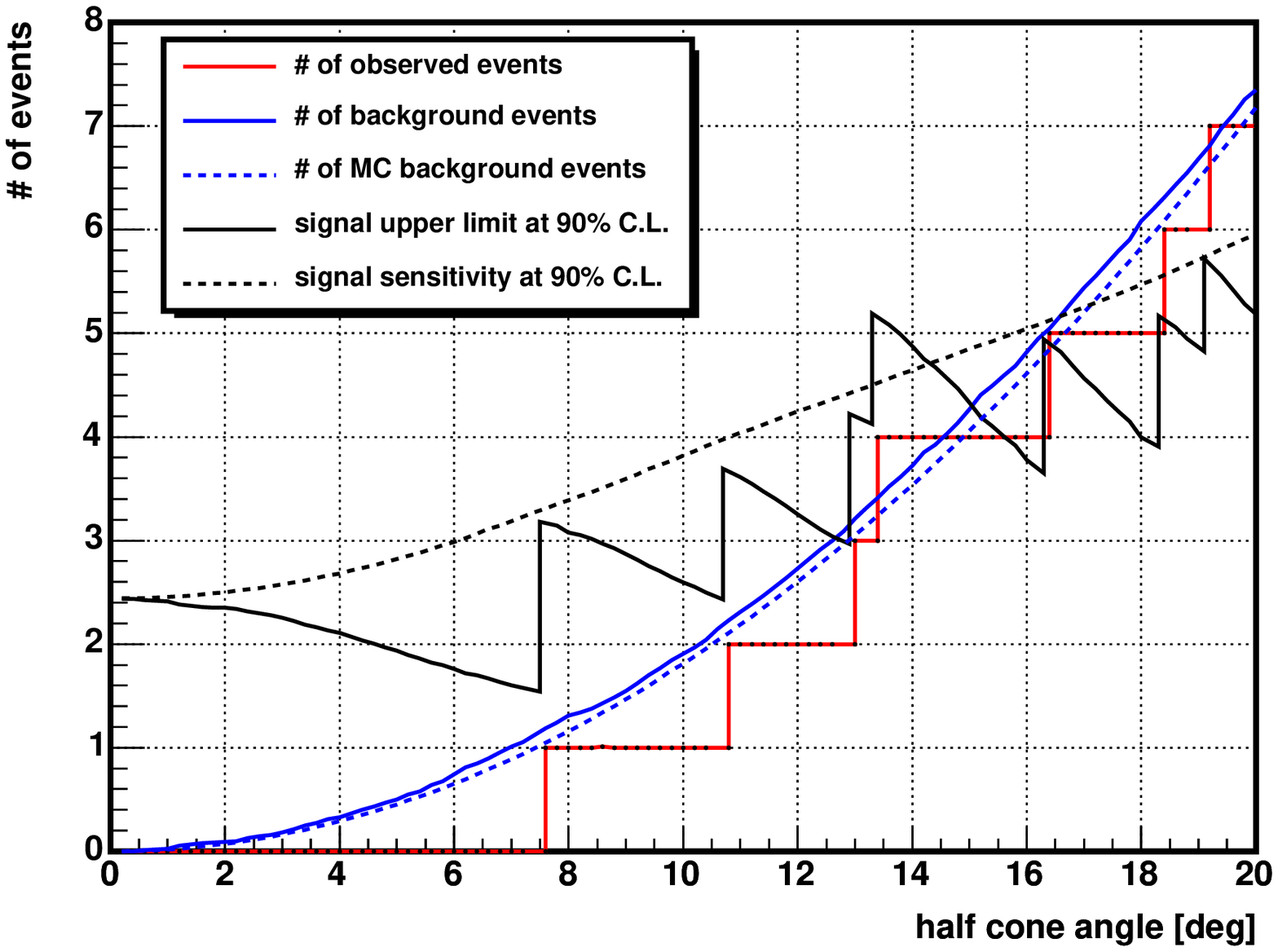,width=8.0cm,clip=}}
\put(7.5, 0.2){\epsfig{file=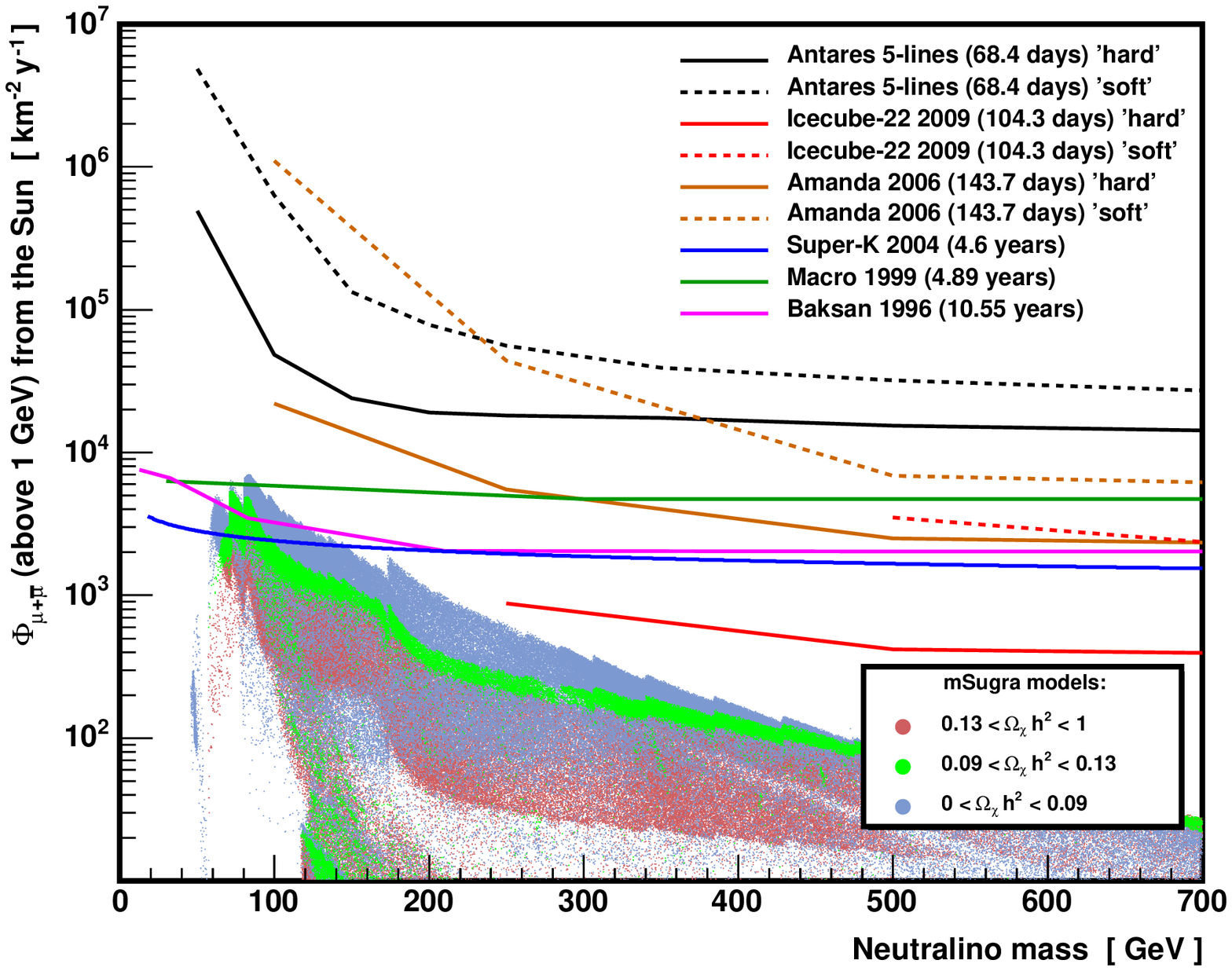,width=7.5cm,clip=}}
 \end{picture}
\caption[]{Left: The number of observed neutrinos and expected background events as a function of the search cone radius around the Sun. Right: The upper limit on the muon flux from the Sun above $E_{\mu} = 1$ GeV as a function of the neutralino mass, in comparison to the expected flux from different mSUGRA models and other indirect detection experiments (IceCube~\cite{ic2}, Amanda~\cite{ama}, Super-K~\cite{super} and Macro~\cite{macro}).}
\label{fig:darkmatter}
\end{figure}

A prediction of the neutrino flux coming from dark matter annihilation in the Sun has been calculated in the 
framework of the mSUGRA model with a neutralino WIMP.
The data taken during 2007 were used to search for possible excess in the 
neutrino flux from the Sun. The livetime of this analysis corresponds to 68.4 days, reduced from the 140 days
due to the condition that the Sun has to be below the horizon.
The number of events in a search cone around the Sun is shown in 
Figure \ref{fig:darkmatter} left as a function of the search cone radius. The number of observed events is in good 
agreement with the expected number of background events 
from Monte Carlo simulation. 
There is no evidence for a flux of neutrinos from the Sun and the upper limits on the corresponding muon flux are
shown in Figure \ref{fig:darkmatter} right and compared with other experiments. 
The mSUGRA parameter space is not reached with this data set, but with a few years of more data ANTARES will become sensitive to 
the focus point region of the mSUGRA parameter space. For further information see~\cite{li}. 

\section{Conclusion}
Antares has been taking data since 2006 with a broad physics program and starts to produce competing results.
Three cosmic neutrino searches (diffuse flux, point sources and dark matter) have been presented. 
No signal has been found in these three analyses; upper limits were set.

\section*{Acknowledgments}
\vspace*{-0.1cm}

I gratefully acknowledge the support of the JAE-Doc postdoctoral programme of CSIC. 
This work has also been supported by the following 
Spanish projects: FPA2009-13983-C02-01, MultiDark Consolider CSD2009-00064, ACI2009-1020 of MICINN and Prometeo/2009/026 of Generalitat Valenciana.

\end{document}